\documentclass[aps,prd,twocolumn,superscriptaddress,longbibliography]{revtex4-2}
\usepackage{bm, amsmath, amsfonts, amssymb, braket}
\usepackage{subfigure}
\usepackage[dvipsnames]{xcolor}
\usepackage{ulem,soul,changes}
\usepackage{dcolumn} % Align table columns on decimal point
\usepackage[colorlinks]{hyperref}
\AtBeginDocument{% optionally set colors to your liking
	\hypersetup{
		urlcolor=black,
		citecolor=green,
		linkcolor=blue,
	}%
}
\usepackage{graphicx}
\usepackage{romannum}
\usepackage{cases,float}
\usepackage[toc,title,page]{appendix}

\usepackage{verbatim}
\usepackage{amstext}
\usepackage{array}
\usepackage{multirow}
\usepackage{physics}
\usepackage{geometry}
\usepackage{enumitem}
\usepackage{indentfirst}

\AtBeginDocument{\pagenumbering{arabic}}

\begin{document}

\title{Mass-independent scheme for enhancing spatial quantum superpositions}
	\newcommand{\affone}{Centre for Quantum Computation and Communication Technology, School of Mathematics and Physics, University of Queensland, Brisbane, Queensland 4072, Australia}
	\newcommand{\afftwo}{Department of Physics and Astronomy, University College London, Gower Street, WC1E 6BT London, United Kingdom.}
	\newcommand{\affthree}{Van Swinderen Institute, University of Groningen, 9747 AG Groningen, The Netherlands.}

	\author{Run Zhou}
	\affiliation{\affthree}
	\author{Ryan J. Marshman}
	\affiliation{\affone}
	\author{Sougato Bose}
	\affiliation{\afftwo}

	\author{Anupam Mazumdar}
	\affiliation{\affthree}
	
	\date{\today}

\begin{abstract}
Placing a large mass in a large spatial superposition, such as a Schr\"odinger Cat state is a significant and important challenge. In particular, the large spatial superposition (${\cal O}(10-100)$ $\mu$m) of mesoscopic masses ($m\sim {\cal O}(10^{-14} -10^{-15})$~kg) makes it possible to test the quantum nature of gravity via entanglement in the laboratory. To date, the proposed methods of achieving this spatial delocalization are to use wavepacket expansions or quantum ancilla (for example spin) dependent forces, all of whose efficacy reduces with mass.  Thus increasing the spatial splitting independent of the mass is an important open challenge. In this paper, we present a method of achieving a mass-independent {\it enhancement} of superposition via diamagnetic repulsion from current-carrying wires.  We analyse an example system which uses the Stern-Gerlach effect to creating a small initial splitting, and then apply our diamagnetic repulsion method to enhance the superposition size ${\cal O}(400-600)$ $\mu$m from an initial modest split of the wavefunction.  We provide an analytic and numeric analysis of our scheme.
\end{abstract}
\maketitle	

%%%%%%%%%%%%%%%%%%%%%%%%%%%%%%%%%%%%%%%%%%%%%%%%%%%%%%%%%%%%%%%%%%%%%%%%%%%%%%%%%%%%%%%%%%%%%%%%
\section{Introduction}

Gravity is special as it is not yet evident whether gravity is a classical or a quantum entity~\cite{Dyson}. It is often thought that any quantum gravitational effects will become important only when we approach the Planck length or the time scale; making it impossible to probe directly. Furthermore, cosmological data, such as perturbations in the cosmic microwave background radiation~\cite{Martin:2017zxs} and the potential B-modes for future detection may not help to settle this outstanding issue~\cite{Ashoorioon:2012kh}. Both astrophysical and cosmological sources contain many uncertainties~\cite{Addazi:2021xuf}.

Despite all these challenges, a tabletop experiment has recently been proposed to explore the quantum origin of gravity with the help of quantum superposition and quantum entanglement in the infrared \cite{bose2017spin,marshman2020locality,bose2022mechanism}, see also~\cite{marletto2017gravitationally,Marletto:2020cdx}. The protocol is known as the quantum gravity induced entanglement of masses (QGEM), which evidences both quantum superposition of geometries \cite{christodoulou2019possibility,christodoulou2022locally}, as well as the exchange of virtual gravitons \cite{marshman2020locality}, that is, spin-2 graviton exchange \cite{bose2022mechanism}), see also \cite{galley2022no}.  Recently, a protocol has been created to entangle the matter with that of the Standard Model photon in a gravitational optomechanical setup~\cite{Biswas:2022qto}, see also~\cite{Carney:2021vvt}. This will probe not only the light bending due to the gravitational interaction but will also probe the spin-2 nature of the graviton mediated entanglement~\cite{Biswas:2022qto}. Note that the entanglement is purely a quantum observable, which measure the quantum correlation in complementary bases, and has no classical analogue whatsoever.
	
One of the key challenges towards realising the QGEM protocol experimentally is to create a large spatial superposition $\delta z\sim {\cal O}(10-100) ~{\rm \mu m}$  for a large mass object ($m\sim {\cal O}(10^{-14}-10^{-15}$ kg), see for details in~\cite{bose2017spin,vandeKamp:2020rqh} and in a free falling setup~\cite{bose2017spin,Toros:2020dbf}. It is also well known that creating a large superposition has many further fundamental applications; one can test  the foundations of quantum mechanics in presence of gravity~\cite{penrose1996gravity,Diosi:1988tf,pearle1989collapse,bassi2013models, nimmrichter2013macroscopicity}, a purely quantum gravitational version of the equivalence principle~\cite{Bose:2022czr},  falsifying spontaneous collapse mechanisms \cite{pearle1989collapse,bassi2013models}, placing a  bound on decoherence mechanisms \cite{PhysRevA.84.052121,romero2011large,Romero_Isart_2010,Tilly:2021qef,Schut:2021svd,Rijavec:2020qxd,carney2018tabletop},  quantum sensors~\cite{Toros:2020dbf,Marshman:2018upe}, probing physics of a fifth fundumental force and the axion~\cite{Barker:2022mdz}, and probing gravitational waves~\cite{Marshman:2018upe}. 

Atom interferometers are well-known to create a large baseline superposition~\cite{Mcguirk2002,Dimopoulos2008a,asenbaum2017phase}, but at masses well below what is required  to test the quantum nature of gravity.  To date, macromolecules represent the heaviest masses placed in a superposition of spatially distinct states \cite{Arndt:1999kyb, gerlich2011quantum}.  There are physical schemes to obtain tiny superpositions of large masses~\cite{Bose:1997mf} or moderate sized ($\sim 10$ nm - $1 \mu$m) superpositions of $\sim 10^{-19} - 10^{-17}$ kg masses \cite{Sekatski_2014,Romero_Isart_2010,romero2011large,Hogan,wan2016free,scala2013matter,yin2013large,clarke2018growingPublished,wood2022spin,kaltenbaek2016macroscopic,pino2018chip,romero2017coherent,kaltenbaek2012MAQRO,arndt:2014testing}.  However,  we require a large spatial superposition of heavy ($m\sim {\cal O}(10^{-17}-10^{-14})$ kg) masses, with the current likely scheme utilising the Stern-Gerlach effect~\cite{Folman2013,folman2019,Folman2018,Margalit2021,Marshman:2021wyk,zhou2022}. In fact, a proof of principle experiment has already been conduced using atoms, showing that such a Stern-Gerlach Interferometer (SGI) for massive objects can indeed be realised~\cite{Margalit2021}. 

The crucial problem now is how to achieve a large spatial separation for these larger mass particles. To the best of our knowledge, there are so far only two realistic types of schemes to separate wave packets, one using wavepacket expansions in conjunction with slits/measurements \cite{Arndt:1999kyb,gerlich2011quantum,romero2011large,kaltenbaek2016macroscopic,pino2018chip,romero2017coherent,kaltenbaek2012MAQRO,arndt:2014testing} and the other using spin-dependent forces \cite{Margalit2021,Marshman:2021wyk,zhou2022}. However, both schemes inevitably become progressively worse as the mass increases. This paper presents a mass-independent experimental scheme to create a large spatial superposition.

There are already some very attractive experimental ideas which have been proposed to create large superposition for masses in the range of $m\sim {\cal O}(10^{-17}-10^{-14})$ kg, see Refs.~\cite{bose2017spin,Marshman:2018upe}, and Refs.~\cite{Pedernales_2020,Marshman:2021wyk,zhou2022}. The latter references considered the effect of induced magnetic potential in the diamond like crystal along with the nitrogen valence (NV)-centred potential. However, it was also noticed that the diamagnetic term in the Hamiltonian inhibits the spatial superposition despite taking into account that the coherence of the NV centre can be maintained for a long enough time~\cite{Pedernales_2020,Marshman:2021wyk}. This was mainly due to the fact that the diamagnetic term creates an harmonic trap and the particle tends to move towards the centre of the potential, therefore inhibiting the growth in the superposition size. This issue was tackled by employing new dynamical techniques such as catapulting the trajectory by assuming non-linear profile for the magnetic field~\cite{zhou2022}. In the current paper, we will explore a yet new possibility of utilising this unavoidable induced diamagnetic effect to enhancing the superposition size. In our analysis, we will use the simple, example model of a magnetic field sourced by current-carrying wires.

Our aim will be to enhance the superposition size using only modest currents, but will require one simple assumption. We will assume that we have already created an initial spatial superposition of the nanocrystal, as could be made using a Stern-Gerlach setup, see~\cite{Pedernales_2020,Marshman:2021wyk,zhou2022}. In particular, we will take for demonstration the initial superposition size $\Delta z_0\sim 1{\rm \mu m}$ and present how this can be increased to a size of $\Delta z\sim {\cal O}(100-500) {\rm \mu m}$. We will also assume that the interferometer setup is one dimensional. And, once we create the superposition, we will assume that the superposition can be closed using the same splitting mechanism, i.e. by the Stern-Gerlach mechanism, see~\cite{zhou2022}. It is for this reason we present this as an enhancement of a spatial superposition.  We will provide  both analytic and numerical analyses. We will take the modest currents $\sim {\cal O}(1)~\text{A}$, and will restrict the example operating time to be within $\sim 0.1$ s while still aiming to maximise the superposition size. 

The maximum size of the superposition will not depend on the mass, thus making this as an attractive scheme for creating macroscopic quantum superposition for any range of masses, provided the initial small superposition is first created. We will show that there are two possible configurations; one where the trajectory is triangular and the other where the trajectory traces its path after interacting with the wire with a closest impact parameter, discussed below.

In our current analysis we will assume that the nanocrystal is not rotating. Recent paper has analysed the rotation of the crystal in Refs.~\cite{PhysRevResearch.3.043174,Stickler_2021,Japha:2022xyg}. We will need revisit this issue separately once we fully analyse the largest superposition size we can obtain. This is similar to the spin coherence problem and will require a dedicated study. We will also not discuss how to close the superposition as it is possible to employ the Stern-Gerlach technique to close the trajectory, see~\cite{zhou2022}. Another generic issue is the phonon vibration causing the decoherence, see~\cite{Henkel:2021wmj}. We will assume that the desired level of internal cooling can be obtained to mitigate the phonon induced decoherence. Given all these challenges can be tackled, we ask how large spatial superposition can we achieve via the diamagnetic contribution.

The paper is organised as follows. In section II, we will discuss the setup with the Hamiltonian. In section III, we will discuss diamagnetic repulsion. In section IV, we will provide the analytical understanding of the interaction of the nanocrystal with the fixed wires. The analysis is very similar to central potential problem in classical mechanics. In section V we will present and discuss our numerical results, and in section VI, we will conclude our paper.

%%%%%%%%%%%%%%%%%%%%%%%%%%%%%%%%%%%%%%%%%%%%%%%%%%%%%%%%%%%%%%%%%%%%%%%%%%%%%%%%%%%%%%%%%%%%%%%%
\section{Setup}

The Hamiltonian of a diamond embedded in a spin in a magnetic field is \cite{Pedernales_2020, Loubser1978}
\begin{equation}\label{Hamitonian}
	\hat{H}=\frac{\hat{\boldsymbol{p}}^{2}}{2 m}+\hbar D \hat{\boldsymbol{S}}^{2}-\frac{\chi_{\rho} m}{2 \mu_{0}} \boldsymbol{B}^{2}-\hat{\boldsymbol{\mu}}\cdot \boldsymbol{B},
\end{equation}	
where $\hat{\boldsymbol{\mu}}=-g\mu_{B}\hat{\boldsymbol{S}}$ is the spin magnetic moment, $g\approx 2$ is the Land\`{e} g-factor, $\mu_{B}=\frac{e \hbar}{2 m_{e}}$ is the Bohr magneton, $\hat{\boldsymbol{S}}$ is spin operator, $\hat{\boldsymbol{p}}$ is momentum operator, $\boldsymbol{B}$ is the magnetic field. $m$, $\hbar$, $D$, $\chi_{\rho}$ and $\mu_{0}$ are scalars representing the mass of the diamond, the reduced Planck constant, NV zero-field splitting, magnetic susceptibility, and vacuum permeability, respectively.
By applying the microwave pulse with an appropriate resonance frequency, the electron spin in the diamond can be coupled with the nuclear spin and the spin magnetic field interaction can be ignored \cite{Shi2014}.  We will  assume that this has happened  prior to the amplification of the superposition size.

Let us first consider the magnetic field generated by a current-carrying wire
\begin{equation}\label{MagneticField}
	\boldsymbol{B}=\frac{ \mu_{0} \vb*{I}\cross \vb*{e_{r}}}{2\pi r},
\end{equation}	
where $\vb*{I}$ is the current carried by a straight wire. $r$ is the vertical distance from the wire to a point in space. $\vb*{e_{r}}$ is the unit vector corresponding to the distance $r$. At this point, the potential energy of the diamond in the magnetic field is given by
\begin{equation}\label{Potential}
	U=-\frac{\chi_{\rho} m}{2 \mu_{0}} \boldsymbol{B}^{2}=\frac{1}{2}m\alpha\frac{I^{2}}{r^{2}},
\end{equation}	
where $\alpha=-\frac{\chi_{\rho} \mu_{0}}{4\pi^{2}}$. Combining Eq.(\ref{Potential}) and Eq.(\ref{MagneticField}), the acceleration of the diamond can be obtained as follows
\begin{equation} \label{Acceleration}
	\vb*{a_{dia}}=-\frac{1}{m}\grad{U}=
		\alpha\frac{I^{2}}{r^{3}}\vb*{e_{r}},
\end{equation}
It can be seen from Eq.(\ref{Acceleration}) that the acceleration caused by the diamagnetic effect (the third term of Eq.(\ref{Hamitonian})) is mass independent. This means we can get a desired acceleration of a diamagnetic object by setting an appropriate distance between the object and the wire and an appropriate current flowing through the wire regardless of the mass of the object. This property provides the possibility of obtaining a large superposition size for massive object.

We will use the trajectory of a classical wave packet to represent the expectation value of the position of the wave packet.

\section{Diamagnetic Repulsion}\label{DiamagneticRepulsion}
The scheme of increasing the superposition size by diamagnetic repulsion is mainly divided into three steps. 

\begin{itemize}

\item
First, we will create a small, initial spatial splitting between the two wave packets, potentially using a Stern-Gerlach apparatus. 

\item Then, a micrwave pulse is used to map the electron spin state to the nuclear spin state, thus allowing us to ignore the spin magnetic field interaction. 

\item Finally, the wave packets enter the magnetic field generated by the current-carrying wires and are further separated under the diamagnetic repulsion, thus achieving a large superposition size. 

\end{itemize}

\begin{figure*}
	%\centering
		\includegraphics[width=0.8\linewidth]{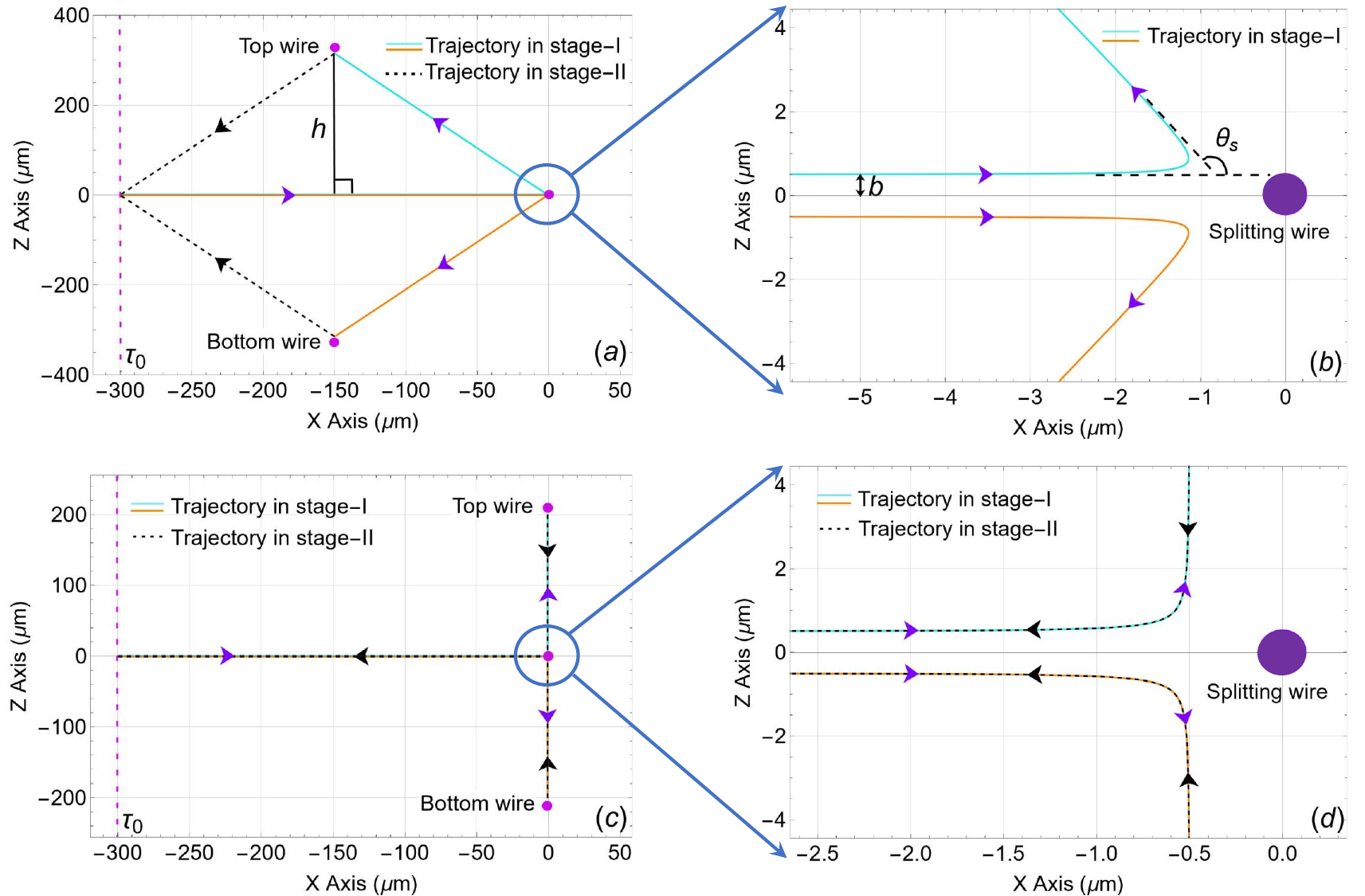}
	
	\caption{ Experimental scheme to create mass independent superposition via diamagnetic repulsion. The wave packet comes out of the Stern-Gerlach interferometer at the moment $\tau_{0}$ with an initial velocity $v_{0}$ parallel to the $x$-axis (polar axis) and enters the magnetic field generated by the current-carrying wire. The purple points (circles) represent straight wires perpendicular to the $x-z$ plane.  The wire at the origin is called the splitting wire. The wire above the $z=0$ axis is called the top wire and the wire below it is called the bottom wire. The blue and orange solid lines represent the wave packet trajectory in stage-$\text{\Romannum{1}}$ and the black dashed lines represent the wave packet trajectory in stage-$\text{\Romannum{2}}$. The purple arrows indicate the direction of motion of the wave packet in stage-$\text{\Romannum{1}}$ and the black arrows indicate the direction of motion of the wave packet in stage-$\text{\Romannum{2}}$. Figs.1(a) and 1(c) show the triangular trajectory and the inverse trajectory respectively, and Figs.1(b) and 1(d) show enlarged images of them close to the splitting wire. The sign $h$ in Fig.1(a) is the distance from the top wire to the bottom edge of the triangle. The sign $b$ in Fig.1(b) is the impact parameter. The initial splitting between two wave packets $\Delta z_{0}=2b$. The sign $\theta_{s}$ is the scattering angle. To simplify the calculation, we set $\tau_{0}=0$.}\label{TrajectoryNumericalResult}
\end{figure*}

The first and second steps are the initial state of the system that we assume. The following analysis is focused on the third step, that is, how to use diamagnetic repulsion to enhance the superposition state. The motion of the wave packet in the magnetic field can be divided into two stages (see Fig.\ref{TrajectoryNumericalResult}).

\begin{itemize}

\item
Stage-\Romannum{1}. The wave packet is incident parallel to the $x$-axis with an initial velocity $v_0$ and its trajectory is deflected by the action of the diamagnetic repulsion. Then the wave packet moves towards the top (bottom) wire until the distance from the x axis reaches a maximum. The spatial distance between the two wave packets also reaches a maximum.

\item
 Stage-\Romannum{2}. The wave packet trajectory is deflected again near the top (bottom) wire and then returns to its initial position. There are two kinds of trajectories that can go back to their initial positions, the triangle trajectory and the inverse trajectory. For triangular trajectory, the wave packet returns to its initial position in the second stage by the shortest path (see Fig.1(a)). For inverse trajectory, the wave packet returns to its initial position in the second stage along the path it took in the first stage (see Fig.1(c)).

\end{itemize}

%%%%%%%%%%%%%%%%%%%%%%%%%%%%%%%%%%%%%%%%%%%%%%%%%%%%%%%%%%%%%%%%%%%%%%%%%%%%%%%%%%%%%%%%%%%%%%%%
%%%%%%%%%%%%%%%%%%%%%%%%%%%%%%%%%%%%%%%%%%%%%%%%%%%%%%%%%%%%%%%%%%%%%%%%%%%%%%%%%%%%%%%%%%%%%%%%	

%%%%%%%%%%%%%%%%%%%%%%%%%%%%%%%%%%%%%%%%%%%%%%%%%%%%%%%%%%%%%%%%%%%%%%%%%%%%%%%%%%%%%%%%%%%%%%%%
\section{Analytic Results}\label{Analyticanalysis} 

In the first stage, the motion of a wave packet in the magnetic field can be viewed as the scattering process of a diamagnetic particle in the magnetic field and then solved analytically using polar coordinates. Since the acceleration of the particle is along the radial direction, the equation of motion can be written as
\begin{equation} \label{EquationOfMotion}
	\dv[2]{r}{t}-r\qty(\dv{\theta}{t})^{2}=\alpha\frac{I^{2}}{r^{3}},
\end{equation}
where the pole is located at the center of the splitting wire and the polar axis is along the $\text{x}$-axis direction. In this case, $\theta$ is the polar angle. The angular momentum of the particle
\begin{align}\label{AngularMomentum} 
	L=|\vb*{L}|
	%&=|m\vb*{r}\cross\vb*{v}|,\nonumber\\
	%&=\qty|m\vb*{r}\cross\qty(\dv{\vb*{r}}{t}+r\dv{\vb*{\theta}}{t})|,\nonumber\\
	%&=\qty|mr\vb*{r}\cross\dv{\vb*{\theta}}{t}|,\nonumber\\
	&=mr^{2}\dv{\theta}{t}.
\end{align}
%Multiply the left side of Eq.(\ref{AngularMomentum}) by 1, we have
%\begin{align}\label{AngularMomentum1}
%	L\times 1=L\dv{\theta}\theta=mr^{2}\dv{t}\theta.
%\end{align}
%Matching the middle and right terms of Eq.(\ref{AngularMomentum1}), we have
%\begin{align}\label{AngularMomentum2}
By using $d/dt=({L}/{mr^{2}})(d/d\theta)$
%\end{align}
and noting that the angular moment is conserved and the initial moment is given by:
\begin{align}\label{AngularMomentum3}
	L=mv_{0}b,
\end{align}
where $v_{0}$ is the initial velocity (parallel to the polar axis direction), $b$ is the vertical distance of the initial position with respect to the polar axis and is called the impact parameter. 
Substituting  Eq.(\ref{AngularMomentum3}) into Eq.(\ref{EquationOfMotion}), we obtain
\begin{align}\label{EquationOfMotion1}
	\dv[2]{u}{\theta}=-ku,
\end{align}
where $u=1/r$, and
\begin{align}\label{Frequency}
	k=1+\alpha\frac{I^{2}}{v_{0}^{2}b^{2}}.
\end{align}
The solution to Eq.(\ref{EquationOfMotion1}) is given by
\begin{align}\label{EquationOfMotion2}
	u=\text{C }\text{cos}(\sqrt{k}\theta-\theta_{0}),
\end{align}
where $\text{C}$ and $\theta_{0}$ are constants determined by the initial conditions. We will solve for $\rm C$ in the Appendix.
Assuming that a particle is incident parallel to the polar axis from infinity and scattered to infinity. When the particle is incident at infinity, $\theta=\pi$, $u=0$, then we have 
\begin{align}\label{SolvingConstant1}
	\sqrt{k}\pi-\theta_{0}=\frac{\pi}{2}.
\end{align}
Rearranging Eq.(\ref{SolvingConstant1}), we get
\begin{align}\label{Constant1}
	\theta_{0}=\qty(\sqrt{k}-\frac{1}{2})\pi.
\end{align}
Here we have taken into account the half period [$-\pi/2,\pi/2$] of the cosine function. When the particle is scattered to infinity, $\theta=\theta_{s}$, $u=0$, then we have
\begin{align}\label{SolvingScatteringAngle}
	\sqrt{k}\theta_{s}-\theta_{0}=-\frac{\pi}{2}.
\end{align}
With a bit of rearrangement of Eq.(\ref{SolvingScatteringAngle}), and by combining Eq.(\ref{Constant1}) we obtain
\begin{align}\label{ScatteringAngle}
	\theta_{s}=\qty(1-\frac{1}{\sqrt{k}})\pi,
\end{align}
which is the scattering angle of the particle. The scattering angle determines the projection of the particle trajectory along the $z$-axis, which is the superposition size we are able to achieve in the experiment. 

Next we study what conditions need to be satisfied for the scattering angle to maximize the superposition size. We assume that the collision between the particle and the wire is an elastic collision. Since the mass of the particle is much smaller than the mass of the wire, which we will also assume is clamped to avoid decoherence, the velocity of the particle before and after the collision can be considered unchanged.
	
 In the experimental scheme, the particle has two elastic collisions with the wire. Since the particle is incident from infinity and then returns to infinity, the time for the particle colliding with the wire and thus the change in velocity can be ignored compared to the total time of motion, and the magnitude of the speed of the particle can be treated as a constant. We consider the particle incident parallel to the $x$-axis with an initial velocity $v_{0}$. The initial position of the particle is $(-x_{0},~b)$ and the value of $x_{0}$ is large enough to allow the particle to be approximated as being infinitely with respect to the magnetic field generated by the wire. 
 
 We set the total time of motion of the particle to be $\tau$. After time $\tau$ the particle returns to its initial position and the trajectory length is $v_{0}\tau$. The trajectory of the particle can be approximated as a triangle with base of length $x_{0}$ (see Fig.1(a)). The problem now becomes where to locate the vertex of the triangle so that the distance $h$ from the vertex to the base takes its maximum value when the sum of the two sides of the triangle is fixed at $v_{0}\tau-x_{0}$. In fact, we are describing the definition of an ellipse. The two focuses of this ellipse are $(-x_{0},~b)$ and $(0,~b)$, and the length of the major axis is $(v_{0}\tau-x_{0})/2$. When the vertex of the triangle is located at the minor axis of this ellipse, the superposition size achieves its maximum value. In this case, the scattering angle $\theta_{s}$ (see Fig.1(b)) of the particle at the splitting wire satisfies
\begin{align}\label{ReflectionAngle}
	\text{cos}\,\theta_{s}=-\frac{x_{0}}{v_{0}\tau-x_{0}}.
\end{align}
The negative sign on the right-hand side of Eq.(\ref{ReflectionAngle}) indicates that the particle is incident from the negative x-axis direction. The length of the minor axis
\begin{align}\label{MinorAxis}
	h&=\frac{v_{0}\tau-x_{0}}{2}\text{sin}\,\theta_{s},\nonumber\\
	&=\frac{v_{0}\tau-x_{0}}{2}\sqrt{1-\text{cos}^{2}\theta_{s}},\nonumber\\
	&=\frac{1}{2}\sqrt{v_{0}^{2}\tau^{2}-2v_{0}x_{0}\tau}.
\end{align}
By combining the symmetry of the incident trajectories of the two wave packets, the largest superposition size for the triangle trajectory can be written as
\begin{align}\label{LargestSuperpositionSize}
	\Delta Z_{maxT}=2h=\sqrt{v_{0}^{2}\tau^{2}-2v_{0}x_{0}\tau}.
\end{align}
Letting the scattering angle in Eq.(\ref{ScatteringAngle}) equal the reflection angle in Eq.(\ref{ReflectionAngle}), combined with Eq.(\ref{Frequency}), we obtain the relationship between the current and the impact parameter and other physical quantities when the superposition size takes its maximum value
\begin{align}\label{CurrentOverImpactParameter}
	\frac{I}{b}=\frac{v_{0}}{\sqrt{\alpha}}\sqrt{\pi^{2}\text{arccos}^{-2}\qty(\frac{x_{0}}{v_{0}\tau-x_{0}})-1}.
\end{align}
Another way to return the particle to its initial position is to follow along its original path. We set the same initial conditions and the velocity as for the triangular trajectory. In the case of inverse trajectory, the superposition size takes a maximum value when the reflection angle of the particle reflected by the splitting wire is $\theta_s =\pi/2$, and then we have
\begin{align}\label{LargestSuperpositionSize1}
	\Delta Z_{maxR}=v_{0}\tau-2x_{0}.
\end{align}
%{\bf why cant I derive it from Eq.(16)?} {\color{blue} Eq.(\ref{MinorAxis}) only holds if the reflection angle satisfies Eq.(\ref{ReflectionAngle}) (triangular trajectory). The reflection angle corresponding to Eq.(\ref{LargestSuperpositionSize1}) is $\pi/2$ (reverse trajectory) and does not satisfy Eq.(\ref{ReflectionAngle}), so we cannot derive Eq.(\ref{LargestSuperpositionSize1}) from Eq.(\ref{MinorAxis}).}

We can see that Eq.(\ref{LargestSuperpositionSize}) and Eq.(\ref{LargestSuperpositionSize1}) do not include the impact parameter $b$. This is because we have assumed that the initial splitting between the wave packets is much less than the superposition size we are going to achieve, so the contribution of the impact parameter to the final superposition size can be ignored. Further, if we are given enough time $\tau$ then provided $v_0\tau\gg x_0\gg b$, both Eq. (\ref{LargestSuperpositionSize} and Eq. (\ref{LargestSuperpositionSize1}) can be approximated in the compact form 
\begin{equation}
	\Delta Z_{maxR}\approx v_{0}\tau
\end{equation}
This simplicity hides the required relationships between scattering angle, impact parameter and current. However, it does highlight clearly that the achievable maximum superposition size is determined by initial conditions and coherence time, not the particle mass.
 
By setting the scattering angle in Eq.(\ref{ScatteringAngle}) equal to $\pi/2$, we can again obtain the relationship between the current and impact parameter and other physical quantities
\begin{align}\label{CurrentOverImpactParameter1}
	\frac{I}{b}=v_{0}\sqrt{\frac{3}{\alpha}}.
\end{align}
To avoid particle collisions with the wire, we need to find the distance of the closest approach $d$ of the particle trajectory to the pole and use this distance as an upper limit on the radius of the wire.

Let us first consider the closest distance $d_{0}$ between the particle and the wire when the particle is incident in the direction of the polar axis. From the conservation of energy, we have
\begin{align}\label{EnergyConservation}
	\frac{1}{2}mv_{0}^{2}=\frac{1}{2}m\alpha\frac{I^{2}}{d_{0}^{2}}.
\end{align}
By rearranging Eq.(\ref{EnergyConservation}), we get
\begin{align}\label{ClosestDistance1}
	d_{0}^{2}=\alpha\frac{I^{2}}{v_{0}^{2}}.
\end{align}

This can then be updated to account for a particle incident {\it parallel} to the polar axis with an impact parameter $b$. When the particle is nearest to the wire, we have the conservation of energy. Here we have only three of the four variables $b,~v, ~d,~ I$ which are independent. The impact parameter $b$ can be determined using the variables $v,~ d, ~I$.
\begin{align}\label{EnergyConservation1}
	\frac{1}{2}mv_{0}^{2}=\frac{1}{2}mv^{2}+\frac{1}{2}m\alpha\frac{I^{2}}{d^{2}},
\end{align}
where $v$ is the velocity of the particle at its nearest point to the wire, $d$ is the distance of the particle trajectory closest to the wire. By combining Eq.(\ref{ClosestDistance1}) and Eq.(\ref{EnergyConservation1}), and simplifying the equation, we obtain
\begin{align}\label{ClosestDistance2}
	\qty(\frac{v}{v_{0}})^{2}=1-\qty(\frac{d_{0}}{d})^{2}.
\end{align}
Using the conservation of momentum, we can find the impact parameter $b$ in terms of $d,~v,~v_0$.
\begin{align}\label{MomentumConservation}
	L=mv_{0}b=mvd,
\end{align}
we have
\begin{align}\label{MomentumConservation1}
	\qty(\frac{v}{v_{0}})^{2}=\qty(\frac{b}{d})^{2}.
\end{align}
Substituting Eq.(\ref{MomentumConservation1}) into Eq.(\ref{ClosestDistance2}), we get
\begin{align}\label{ClosestDistance3}
	d=\sqrt{b^2+d_{0}^{2}}.
\end{align}
Eq.(\ref{ClosestDistance3}) is consistent with our intuition that the nearest distance between the particle and the wire increases with increasing impact parameter and current, and decreases with increasing initial velocity. As the initial velocity approaches infinity, $d_{0}$ approaches 0 and the distance $d$ approaches the impact parameter $b$.

%%%%%%%%%%%%%%%%%%%%%%%%%%%%%%%%%%%%%%%%%%%%%%%%%%%%%%%%%%%%%%%%%%%%%%%%%%%%%%%%%%%%%%%%%%%%%%%%
\section{Numerical results}\label{NumericalResults} 

In this section, we will solve the wave packet trajectory numerically using the equation of motion Eq.(\ref{Acceleration}). We will take into account that the initial conditions satisfy Eq.(\ref{CurrentOverImpactParameter}), or Eq.(\ref{CurrentOverImpactParameter1}) to obtain the triangular trajectory, or the inverse trajectory. All the three wires are switched on during the movement of the wave packets.

We have assumed the initial spatial splitting between the two wave packets $\Delta z_{0}=1$ ${\rm\mu  m}$ ($b=0.5\,\mu \text{m}$), and the initial velocity $v_{0}=0.01{\rm m/s}$, which is parallel to the $x$-axis, as shown in Figs.~(1(a),1(c)).

 Both of these initial conditions could potentially be achieved in a nanocrystal with a mass of $10^{-15} {\rm kg}$ by using the Stern-Gerlach apparatus with a magnetic field gradient of $10^{4}{\rm T/m}$ \cite{zhou2022}. In this paper, we will focus on how to use the diamagnetic repulsion to further increase the spatial separation between wave packets. 
 %In this paper we do not study the process of preparing the initial states.

%%%%%%%%%%%%%%%%%%%%%%%%%%%%%%%%%%%

\subsection{Numerical result for triangular trajectory}

We set the current through the splitting wire according to Eq.(\ref{CurrentOverImpactParameter}) and then place the top and bottom wires at the minor axis of the ellipse and adjust the current through the wire so that the trajectory of the wave packet forms a triangle back to its initial position. 

The numerical result of the triangular trajectory is shown in Fig.1(a). The wave packets incident from the initial position $(-300,\pm 0.5)$, and the motion time $\tau=0.1$ $\text{s}$. The coordinates of the top and bottom wires are $(-150,\pm 316.5)$. The current through the splitting wire is 0.925273 A and the current through the top (bottom) wire is 1.57 A. In this case we obtain a maximum superposition size of around $628$ ${\rm \mu m}$. Substituting these initial conditions into Eq.(\ref{LargestSuperpositionSize}), gives a superposition size of around $632~{\rm \mu m}$. These two results are close. This indicates that our assumption of a constant motion of the wave packet and a negligible contribution of the impact parameter $b$ to the final superposition size is reasonable.

The closest distance from the triangular trajectory to the splitting wire is $1.39266$ ${\rm \mu m}$ and the closest distance to the top (bottom) wire is $2.29929$ ${\rm \mu m}$. By using Eq.(\ref{ClosestDistance3}), we can calculate the closest distance from the wave packet trajectory to the splitting wire to be 
$1.39269 $ microns, which agrees well with the numerical results. 

If the minimum distance from the wave packet trajectory to the wire is considered to be the maximum radius of the wire, the minimum current density $\rho_{current}$ required for the experiment is around $0.15$ $\text{A}$ $\mu \text{m}^{-2}$. In the laboratory, current densities of $\order{10^{0}}\sim\order{10^{1}}$ $\text{A}$ $\mu \text{m}^{-2}$ can be achieved using carbon nanotubes and graphene \cite{Yao2000,Wei2001,Murali2009}. The magnetic field experienced by the wave packet when the triangular trajectory is closest to the wire is around $0.13$ $\text{T}$.

\subsection{Numerical results for inverse trajectory}

The numerical analysis of the inverse trajectory is almost the same as that of the triangular trajectory section. Here we set the current through the splitting wire according to Eq.(\ref{CurrentOverImpactParameter1}), then place the top and bottom wires on the $x=-0.5$ $\mu \text{m}$ axis and adjust the current through the wire so that the trajectory of the wave packet returns to its initial position along its original path.

The numerical result of the inverse trajectory is shown in Fig.1(c). The wave packets also incident from the initial position $(-300,\pm 0.5)$, and the motion time $\tau=0.1$ $\text{s}$. The coordinates of the top and bottom wires are $(-0.5,\pm 200)$. The current through the splitting wire is 0.616467 A and the current through the top (bottom) wire is 0.00823 A. In this case we obtain a maximum superposition size of $399.977$ $\mu \text{m}$, which is very close to the result of $400$ $\mu \text{m}$ calculated by Eq.(\ref{LargestSuperpositionSize1}). The closest distances from the inverse trajectory to the splitting wire is $0.999996$ ${\rm \mu m}$ and the closest distance to the top (bottom) wire is $0.0115617$ ${\rm \mu m}$. Following the same discussion as for the triangular trajectory, the minimum current density required for the inverse trajectory scheme is around $19.6$ $\text{A}$ $\mu \text{m}^{-2}$. The magnetic field experienced by the wave packet when the inverse trajectory is closest to the wire is around $0.14$ $\text{T}$.

Compared to the triangular trajectory, the superposition size achieved by the inverse trajectory is smaller ($400$ $\mu \text{m}$ $vs.$ $628$ $\mu \text{m}$ in our case) with the same motion time. The advantage of the inverse trajectory is that when the wave packet returns to its initial position, the velocity of the wave packet is of the same magnitude as that of the initial velocity and in the opposite direction. Which means that we can close the trajectories of the wave packets by using an inverse process of creating an initial splitting between the wave packets, for example, via the Stern-Gerlach apparatus. This also helps recover the spin coherence \cite{zhou2022}.

\subsection{Scaling of superposition size}
\begin{figure}\label{Scaling}
		\includegraphics[width=0.9\linewidth]{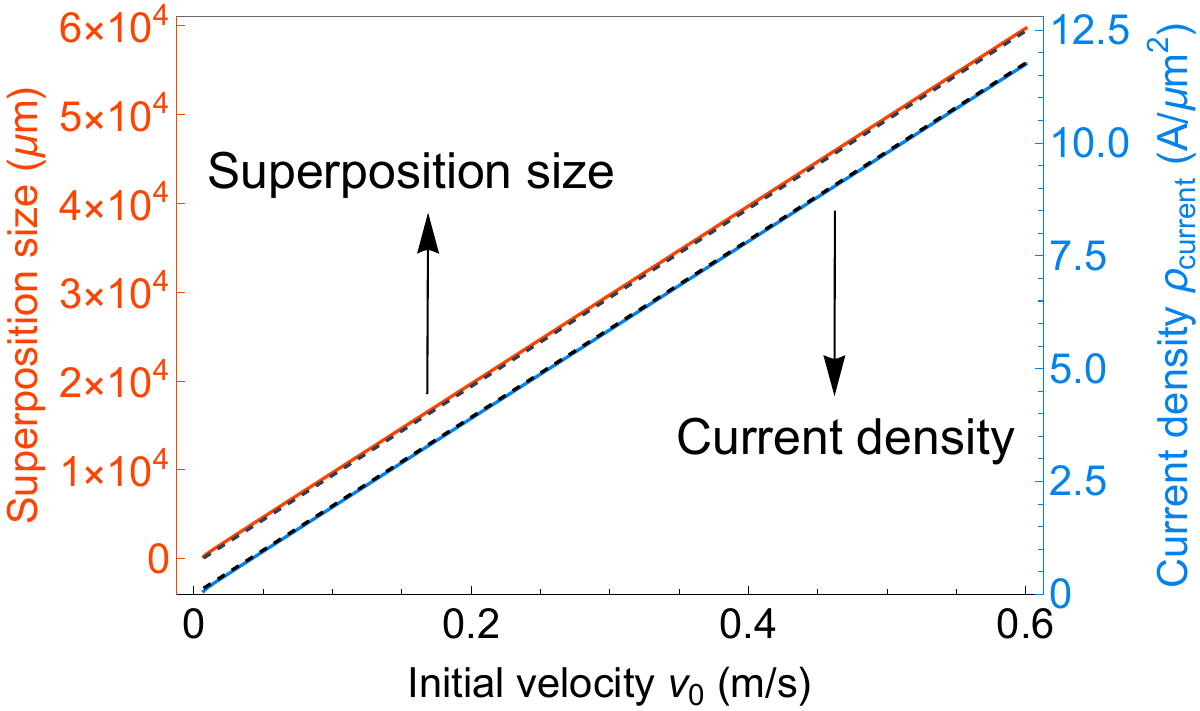}

	\caption{We have shown here the size of the superposition and the required current density with regards to the initial velocity. The upper red solid line and the black dashed line correspond to the superposition size that can be achieved for the triangular trajectory and the inverse trajectory for different initial velocities, respectively. The lower blue solid line and the black dashed line correspond to the current density required to form a triangular trajectory and an inverse trajectory for different initial velocities, respectively. Here we have set the impact parameter: $b=0.5~{\rm \mu m}$, the initial coordinate: $-x_{0}=-300~{\rm \mu m}$, and the motion time: $\tau=0.1~{\rm s}$.} \label{SSandCD-VS-InitialVelocity}
\end{figure}

The most important factor affecting the superposition size is the initial velocity of the wave packet with respect to the wire. According to Eq.(\ref{LargestSuperpositionSize}) and Eq.(\ref{LargestSuperpositionSize1}), we can obtain the change in superposition size with initial velocity for the triangular trajectory and the inverse trajectory, as shown in Fig.\ref{SSandCD-VS-InitialVelocity}. The superposition sizes that can be achieved for both trajectories are of the same order of magnitude and increase linearly with initial velocity. In the other hand, the larger initial velocity means that we need a larger current density to deflect the trajectory to form a triangular or inverse trajectory. 

Combining Eq.(\ref{ClosestDistance1}) and (\ref{ClosestDistance3}) gives the current density
\begin{align}\label{CurrentDensity}
	\rho_{current}=\frac{I}{\pi d^{2}}=\frac{1}{b}\frac{C_{1,2}v_{0}^{2}}{\pi(v_{0}^{2}+\alpha C_{1,2}^{2})},
\end{align}
where $C_{1}=I/b$ corresponding to triangular trajectory (see Eq.(\ref{CurrentOverImpactParameter})), and $C_{2}=I/b$ corresponding to inverse trajectory (see Eq.(\ref{CurrentOverImpactParameter1})). Fig.\ref{SSandCD-VS-InitialVelocity} also shows the change in current density with initial velocity. We can see that when the initial velocity is large enough, the current density corresponding to the triangular trajectory and the inverse trajectory changes linearly with the initial velocity and the curves almost coincide. If we consider that the initial velocity and motion time of the wave packet are determined, then $C_{1}$ and $C_{2}$ are constants and the current density is inversely proportional to the initial splitting between wave packets. That is, for smaller initial splitting, we need larger current density to achieve the same superposition size. 

%%%%%%%%%%%%%%%%%%%%%%%%%%%%%%%%%%%%%%%%%%%%%%%%%%%%%%%%%%%%%%%%%%%%%%%%%%%%%%%%%%%%%%%%%%%%%%%%
%%%%%%%%%%%%%%%%%%%%%%%%%%%%%%%%%%%%%%%%%%%%%%%%%%%%%%%%%%%%%%%%%%%%%%%%%%%%%%%%%%%%%%%%%%%%%%%%

%%%%%%%%%%%%%%%%%%%%%%%%%%%%%%%%%%%%%%%%%%%%%%%%%%%%%%%%%%%%%%%%%%%%%%%%%%%%%%%%%%%%%%%%%%%%%%%%
\subsection{Comparing schemes for creating spatial superposition}

 Here we briefly compare the different schemes for creating the spatial superposition. The larger the mass and size of the superposition state and the longer the coherence time, the more likely it is to induce decoherence due to thermal radiation, gas collisions and other noise (e.g. fluctuations in electromagnetic fields, phonon excitations, etc.) \cite{bose2017spin,marshman2020locality,vandeKamp:2020rqh,PhysRevA.84.052121,romero2011large,Romero_Isart_2010,gonzalez2021,HornbergerEtAl2004,Sinha:2022snc}. Different combinations of (a) small mass ($\sim 10^{-25}\, {\rm kg}$, Rb-87), large superposition size ($\sim 54\,{\rm cm}$) and short coherence time ($\sim 2\,{\rm s}$) \cite{kovachy2015}, and (b) small mass ($\sim 10^{-25}\, {\rm kg}$, Cesium-133), small superposition size ($\sim 4\,{\rm \mu m}$) and long coherence time ($\sim 20\,{\rm s}$) \cite{Xu:2019vlt}, as well as (c) large mass ($\sim 10^{-23}\, {\rm kg}$, Oligoporphyrins), small superposition size ($\sim 100\,{\rm nm}$) and short coherence time ($\sim 8\,{\rm ms}$) \cite{FeinEtAl2019} have been implemented in the laboratory recently. However, the methods of light-pulse and matter-wave diffraction used in schemes (a) to (c) are difficult to apply to larger masses of superposition states (e.g. $10^{-17}-10^{-14}$ kg) on account of internal heating and too short de Broglie wavelengths \cite{gonzalez2021}. In Ref. \cite{Margalit2021} a full-loop Stern-Gerlach interferometer was realized using magnetic field and spin coupling for a mass of $\sim 10^{-25}\, {\rm kg}$ (Rb-87), a superposition size of $\sim 2\,{\rm \mu m}$ and a coherence time of $\sim 4\,{\rm ms}$. The advantage of this method is that it can potentially be used to create large spatial superposition ($1-10\,{\rm \mu m}$) with large masses ($10^{-17}-10^{-14}$ kg) \cite{Margalit2021,Marshman:2021wyk,zhou2022}. But since the effect of using magnetic field and spin coupling to separate wave packets is inversely proportional to mass, this limits the superposition size that can be achieved for macroscopic superposition. Our solution is to increase a small superposition size (e.g. 1 ${\rm \mu m}$) to the order of $ 10^{2} {\rm \mu m}$ by mass independent diamagnetic repulsion effects (Eq.(\ref{Acceleration})). We note that recently there have been some novel ideas to enhance entanglement signals as well, for example by coupling small mass two-level system to a massive mediator oscillator to amplify the entanglement phase \cite{Pedernales:2021dja} or by amplified-weak-value phenomenon in two-setting Einstein-Podolsky-Rosen steering to increase the sensitivity to weak entanglement signals \cite{Feng:2022hfv}. Interestingly, in principle, our scheme can be combined with those in \cite{Pedernales:2021dja, Feng:2022hfv} to significantly improve the sensitivity of the detection of entangled signals.

% There are no fundamental limitations to the creation of the macroscopic superposition state, the difficulty lies in the cooling of the macroscopic object to the ground state and in maintaining the high vacuum environment.

%%%%%%%%%%%%%%%%%%%%%%%%%%%%%%%%%%%%%%%%%%%%%%%%%%%%%%%%%%%%%%%%%%%%%%%%%%%%%%%%%%%%%%%%%%%%%%%%
\section{Conclusion and Discussion}

In this paper, we propose a mass independent scheme to obtain large spatial superposition. We analytically and numerically studied two possible trajectories which we call the triangular trajectory and the inverse trajectory. We found that given an initial spatial splitting $\Delta z_{0}=1$ $\mu \text{m}$, an initial velocity $v_{0}=0.01$ $\text{m}/\text{s}$ and total time $\tau=0.1$ $\text{s}$, the maximum superposition size achievable with the triangular trajectory is $632~{\rm \mu m}$, and $400$ $\mu \text{m}$ for the inverse trajectory. This represents a 400 to 632 times enhancement on the initial splitting and is completely independent of the mass used.

For both the triangular and the inverse trajectories, we chose to require the wave packets to return approximately to their initial positions. In principle, however, as long as the spatial separation between the wave packets is less than or equal to the initial splitting, we can likely recombine the trajectories of the two wave packets coherently via the Stern-Gerlach setup, see~\cite{zhou2022}. 

 It may not be necessary to accelerate the nanocrystal to the desired initial velocity by the Stern-Gerlach apparatus. We could for example consider the nanocrystal falling in a gravitational field while the wires remain stationary in the gravitational field. Since the gravitational acceleration is about 9.8 ${\rm m/s^{2}}$, we can easily achieve a large relative velocity between the nanocrystal and the wire and thus realize the same scattering process. We will provide the details of this construction in near future.

In addition, further considerations will be required, depending on the method of generating the initial splitting and final interference. For example, a high-precision controlled magnetic field is needed to finally recover spin coherence \cite{Schwinger1988} if using a Stern-Gerlach interferometer. While not considered in great detail here, this spin coherence has been studied elsewhere \cite{Bar-Gill2013, Abobeih2018, Muhonen2014, Farfurnik2015, Knowles2014}. Indeed spin coherence times have been gradually improving  (approaching 1 second \cite{Bar-Gill2013, Abobeih2018}, even 30 seconds \cite{Muhonen2014, Farfurnik2015}) adapting these to nanocrystals remains an open problem, but there is no fundamental constraints \cite{Knowles2014}. The spatial coherence times can be made 100 seconds, see \cite{bose2017spin, Toros:2020dbf, vandeKamp:2020rqh}. There are other very important challenges, achieving pressures, temperatures, distances from other sources and fluctuations \cite{vandeKamp:2020rqh}.  In addition to these, we will need to ensure that 
we can close the interferometer loop \cite{Schwinger1988}. This requires the position, the momentum and the spin direction to be aligned, see \cite{Japha:2022xyg}. Furthermore, we will need to analysis the spin coherence including the rotation of the nanocrystal if the NV centre is not at the centre of the nanocrystal \cite{Japha:2022xyg} and the excitations of the phonons~\cite{Henkel:2021wmj}. However, such considerations are left for further study. Indeed, given this diamagnetic enhancement does not itself require any spin-based manipulation, there may be other, simpler candidate systems for creating (and closing) the initial small spatial splitting which does not concern itself with such issues. Although it will likely require similar efforts to ensure coherent interference is possible.

\begin{acknowledgements}
	We would like to thank Ron Folman for helpful discussions.
	R. Z. is supported by China Scholarship Council (CSC) fellowship. R. J. M. is supported by the Australian Research Council (ARC) under the Centre of Excellence for Quantum Computation and Communication Technology (CE170100012). 
	AM would like to thank Institute for Advanced Study Princeton for their kind hospitality and for hosting the author during the completion of this paper. 
	%AM's research is funded by the Netherlands Organisation for Science and Research (NWO) grant number 680-91-119. SB would like to acknowledge EPSRC Grant Nos. EP/N031105/1 and EP/S000267/1.
\end{acknowledgements}

\bibliography{references.bib} 
%%%%%%%%%%%%%%%%%%%%%%%%%%%%%%%%%%%%%%%%%%%%%%%%%%%%%%%%%%%%%%%%%%%%%%%%%%%%%%%%%%%%%%%%%%%%%%%%

\begin{appendices}
\section{Solving amplitude C}

To solve the constant $\text{C}$ in Eq.(\ref{EquationOfMotion2}), we consider a point on the incident trajectory which is at a distance $\text{b}$ from the polar axis and satisfies $\sqrt{k}bu\ll 1$. Then we have
\begin{align}\label{SolvingConstant2}
	u&=\text{C}\text{cos}\qty(\sqrt{k}(\pi-\text{arcsin}(bu))-\theta_{0}),\nonumber\\
	&\approx\text{C}\text{cos}\qty(\sqrt{k}(\pi-bu)-\theta_{0}),\nonumber\\
	&=\text{C}\text{cos}(\sqrt{k}\pi-\theta_{0})\text{cos}(\sqrt{k}bu)\nonumber\\
	&\quad+\text{C}\text{sin}(\sqrt{k}\pi-\theta_{0})\text{sin}(\sqrt{k}bu),\nonumber\\
	&=\text{C}\text{sin}(\sqrt{k}bu),\nonumber\\
	&\approx\text{C}\sqrt{k}bu,
\end{align}
where the relation $\sqrt{k}\pi-\theta_{0}=\pi/2$ is used. The constant $\text{C}$ can be obtained by solving Eq.(\ref{SolvingConstant2}), which is
\begin{align}\label{Constant2}
	\text{C}=\frac{1}{\sqrt{k}b}.
\end{align}
%%%%%%%%%%%%%%%%%%%%%%%%%%%%%%%%%%%%%%%%%%%%%%%%%%%%%%%%%%%%%%%%%%%%%%%%%%%%%%%%%%%%%%%%%%%%%%%%
\begin{figure}[t]
	\includegraphics[width=0.9\linewidth]{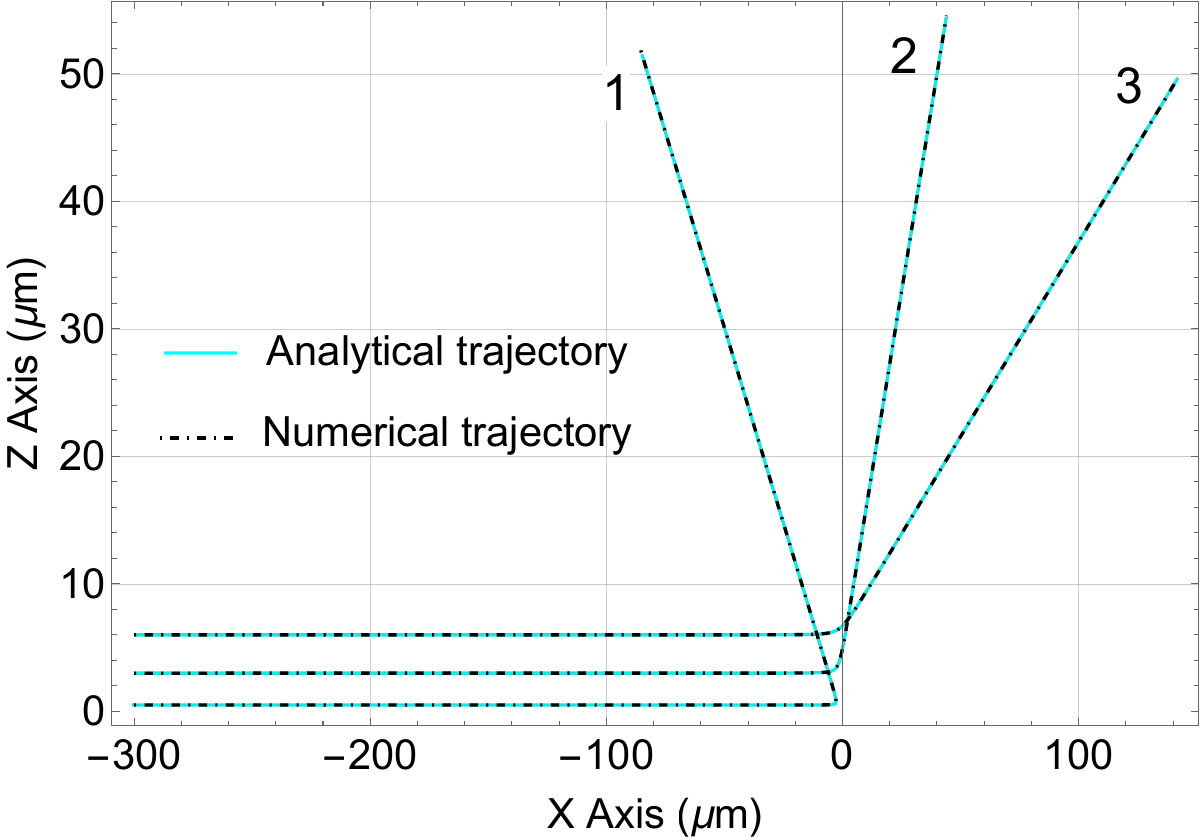}
	
	\caption{We have shown the comparison of the analytical and the numerical trajectories of the particle in the magnetic field. The solid blue line is the analytical trajectory and the black dot-dashed line is the numerical trajectory. Trajectories 1, 2 and 3 correspond to impact parameters of 0.5 $\mu\text{m}$, 3 $\mu\text{m}$ and 6 $\mu\text{m}$ respectively. These three trajectories correspond to particles with an initial velocity of 0.01 $\text{m}/s$. The splitting wire is located at the origin and the current through the wire is 2 $\text{A}$.}\label{AnalyticalAndNumericalTrajectory}
\end{figure}
%%%%%%%%%%%%%%%%%%%%%%%%%%%%%%%%%%%%%%%%%%%%%%%%%%%%%%%%%%%%%%%%%%%%%%%%%%%%%%%%%%%%%%%%%%%%%%%%

Combining Eq.(\ref{EquationOfMotion1}), Eq.(\ref{Constant1}) and Eq.(\ref{Constant2}), we can finally obtain the analytical trajectory of the particle as
\begin{align}\label{AnalyticalTrajectory}
	r=\frac{\sqrt{k}b}{\text{cos}\qty(\sqrt{k}\theta-\qty(\sqrt{k}-\frac{1}{2})\pi)}.
\end{align}

We can solve the trajectory of the particle numerically using Eq.(\ref{Acceleration}) and then compare it with the analytical solution in Eq.(\ref{AnalyticalTrajectory}), as shown in Fig.\ref{AnalyticalAndNumericalTrajectory}. The analytical results fit well with the numerical results.

%%%%%%%%%%%%%%%%%%%%%%%%%%%%%%%%%%%%%%%%%%%%%%%%%%%%%%%%%%%%%%%%%%%%%%%%%%%%%%%%%%%%%%%%%%%%%%%%

\section{Uncertainties in the parameters}
Here we briefly discuss the effect of uncertainties in initial position and velocity on the accuracy of the final position. Using the expression for the scattering angle, Eq.(\ref{ScatteringAngle}), to differentiate the impact parameter and initial velocity gives
\begin{align}\label{AngleFluctuation1}
	\Delta \theta_{si}=-A\frac{\Delta b_{0}}{b},
\end{align}
and
\begin{align}\label{AngleFluctuation2}
	\Delta \theta_{sv}=-A\frac{\Delta v_{0}}{v_{0}},
\end{align}
where
\begin{align}
	A=\frac{k-1}{k^{\frac{3}{2}}}\pi.
\end{align}
The expression for $k$ is given by Eq.(\ref{Frequency}). 

%%%%%%%%%%%%%%%%%%%%%%%%%%%%%%%%%%%%%%%%%%%%%%%%%%%%%%%%%%%%%%%%%%%%%%%%%%%%%%%%%%%%%%%%%%%%%%%%%%%%%%%%%%%%%%
\begin{figure*}[t]
	\centering
	\includegraphics[width=0.34\linewidth]{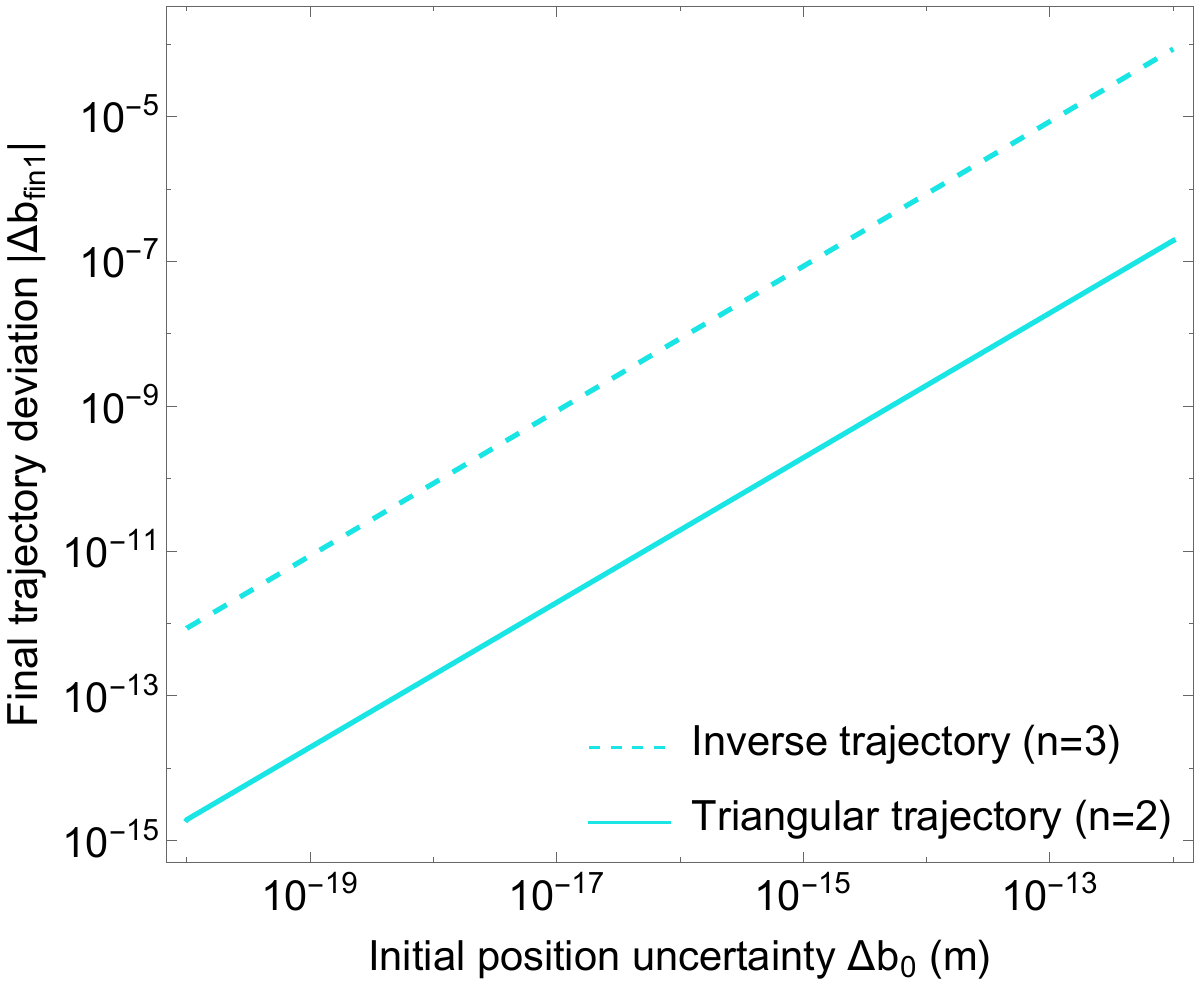}\hspace{0.5cm}
	\includegraphics[width=0.34\linewidth]{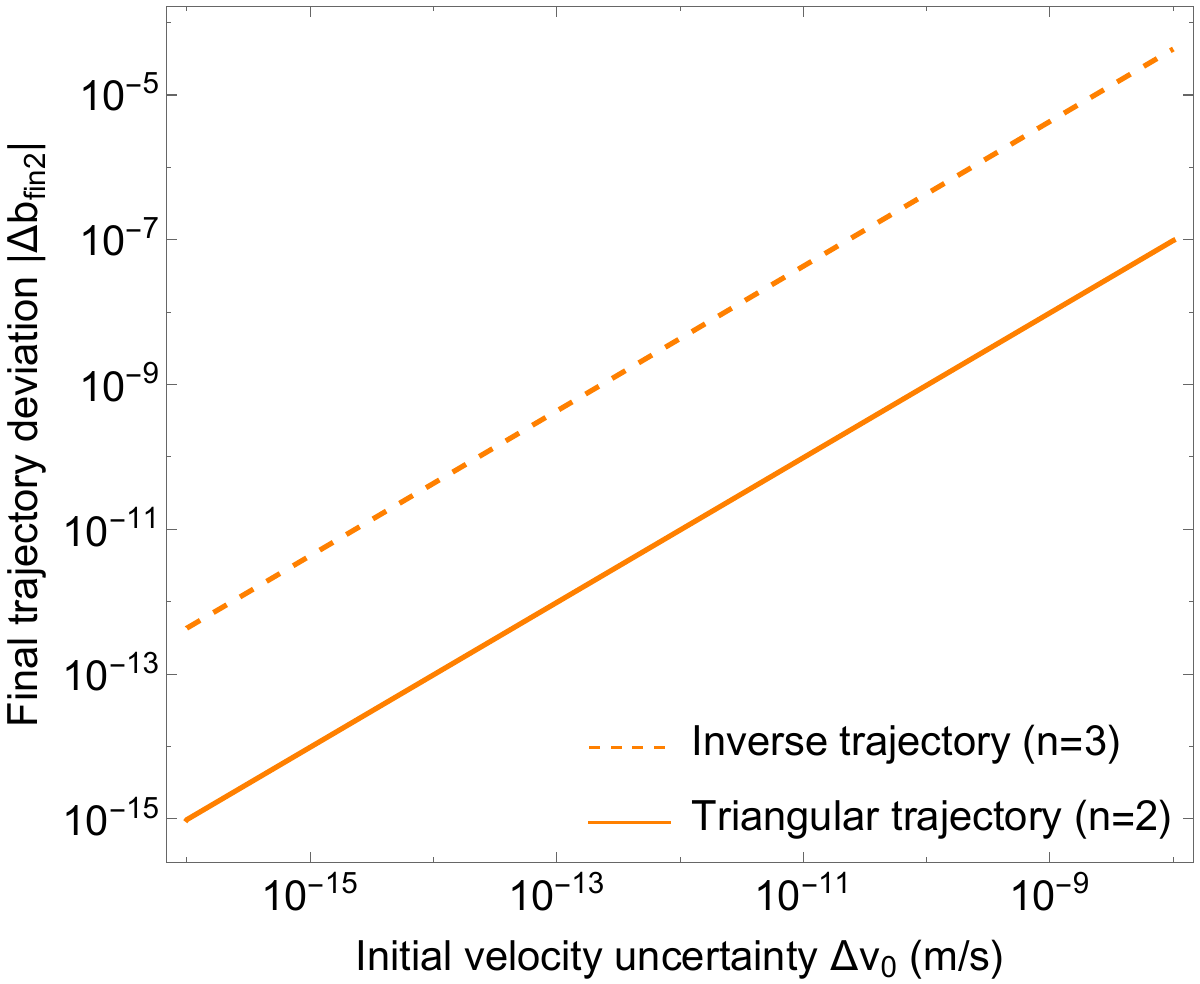}
	\caption{We have shown here the relation between the deviation in the final trajectory, and the uncertainty of the initial position and velocity. The blue and the orange dashed lines correspond to the inverse trajectories. The solid blue and the orange lines correspond to the triangular trajectories. We have set $\theta_s=3\pi/4, b=0.5\,{\rm \mu m}, L=300\,{\rm \mu m}$.}\label{Deviation}
\end{figure*}
%%%%%%%%%%%%%%%%%%%%%%%%%%%%%%%%%%%%%%%%%%%%%%%%%%%%%%%%%%%%%%%%%%%%%%%%%%%%%%%%%%%%%%%%%%%%%%%%%%%%%%%%%%%%%%

To simplify the discussion, we assume that the impact parameter is $b$ and scattering angle is $3\pi/4$ for each scattering, and that the distance from the initial position to the wire and from the wire to the wire is $L$. With the help of Eq.(\ref{AngleFluctuation1}) and (\ref{AngleFluctuation2}) we can solve the deviation in the trajectories, when the wave packet returns to its initial position:
\begin{align}\label{Deviation1}
	\abs{\Delta b_{fin1}}=A^{n}\frac{\Delta b_{0}}{b^{n}}L^{n},
\end{align}
and
\begin{align}\label{Deviation2}
	\abs{\Delta b_{fin2}}=A^{n}\frac{\Delta v_{0}}{v_{0}b^{n-1}}L^{n}.
\end{align}
$\Delta b_{fin1}$ and $\Delta b_{fin2}$ are trajectory deviations caused by inaccuracies in the initial position and initial velocity, respectively. The superscript $``n"$ represents the number of scatterings between the diamond and the wire, $n=2$ for the triangular trajectory, and $n=3$ is for the inverse trajectory. The final trajectory deviation with the initial position and the velocity uncertainty is shown in Fig.\ref{Deviation}. For the same initial position, or velocity uncertainty, the deviation of the inverse trajectory at the final position is approximately three orders of magnitude greater than the deviation of the triangular trajectory (we have fixed $L=300\,{\rm \mu m}$).

It is worth noting that the uncertainty in the initial position and the velocity here is not the width of the position and momentum of the wave packet. The uncertainty here refers to the environmental perturbation of the wave packet position and momentum, which can be much smaller than the position and momentum width of the wave packet. For example, in 2018 Margalit et al. used Stern-Gerlach interferometry to separate $^{87}{\rm Rb}$ atoms and eventually recover spin coherence by precisely controlling the current and pulse timing \cite{Folman2018}. The condition for recovering spin coherence is that the classical uncertainties in position and momentum are much smaller than the quantum uncertainties \cite{Schwinger1988}.

\end{appendices}

\end{document}